\documentclass[twocolumn,tighten]{aastex62}
\usepackage{natbib}

\usepackage[hang,raggedright,tight]{subfigure}
\usepackage{longtable}
\usepackage[figuresright]{rotating}
\usepackage{float}
\usepackage[toc,page]{appendix}

\usepackage{gensymb}
\usepackage{threeparttable}
\usepackage{diagbox}
\usepackage{pbox}
\usepackage{multirow}

\shorttitle{decay-rate }
\shortauthors{Dong et al.}

\begin{document}
\title{The first direct measurement of gravitational potential decay rate at cosmological scales and improved dark energy constraint}
\author[0000-0003-0296-0841]{Fuyu Dong}
\altaffiliation{dongfy2020@kias.re.kr}
\affiliation{School of Physics, Korea Institute for Advanced Study (KIAS), 85 Hoegiro, Dongdaemun-gu, Seoul, 02455, Republic of Korea}
\author{Pengjie Zhang}
\altaffiliation{zhangpj@sjtu.edu.cn}
\affiliation{Department of Astronomy, School of Physics and Astronomy, Shanghai Jiao Tong University, Shanghai, 200240, China} 
\affiliation{Division of Astronomy and Astrophysics, Tsung-Dao Lee Institute, Shanghai Jiao Tong University, Shanghai, 200240, China}
\affiliation{Key Laboratory for Particle Astrophysics and Cosmology
(MOE)/Shanghai Key Laboratory for Particle Physics and Cosmology,
China} 
\author[0000-0001-8771-306X]{Zeyang Sun}
\affiliation{Department of Astronomy, School of Physics and Astronomy, Shanghai Jiao Tong University, Shanghai, 200240, China}
\affiliation{Key Laboratory for Particle Astrophysics and Cosmology
(MOE)/Shanghai Key Laboratory for Particle Physics and Cosmology,
China} 
\author{Changbom Park}
\affiliation{School of Physics, Korea Institute for Advanced Study (KIAS), 85 Hoegiro, Dongdaemun-gu, Seoul, 02455, Republic of Korea}
\begin{abstract}
The integrated Sachs-Wolfe (ISW) effect probes the decay rate ($DR$) of large scale gravitational potential and therefore provides unique constraint on dark energy (DE). However its constraining power is degraded by the ISW measurement, which relies on cross-correlating with the large scale structure (LSS) and suffers from uncertainties in galaxy bias and matter clustering. In combination with lensing-LSS cross-correlation, $DR$ can be isolated in a way free of uncertainties in galaxy bias and matter clustering. We applied this proposal to the  combination of the DR8 galaxy catalogue of DESI imaging surveys and Planck cosmic microwave background (CMB) maps. We achieved the first  $DR$ measurement, with a total significance of $3.2\sigma$. We verified the measurements at three redshift bins ($[0.2,0.4)$, $[0.4, 0.6)$, $[0.6,0.8]$), with two LSS tracers (the "low-density points" and the conventional galaxy positions). Despite its relatively low S/N, the addition of $DR$ significantly improves dark energy constraints, over SDSS baryon acoustic  oscillation (BAO) data alone or Pantheon supernovae (SN) compilation alone. For flat $w$CDM cosmology, the improvement in the precision of $\Omega_m$ is a factor of 1.8 over BAO  and 1.5 over SN. For the DE equation of state $w$, the improvement factor is 1.3 over BAO  and 1.4 over SN. These improvements demonstrate $DR$ as a useful cosmological probe, and therefore we advocate its usage in future cosmological analysis. 
\end{abstract}

\section{Introduction}
A variety of observable effects are induced by the interactions between CMB photons and the large-scale structure (LSS) that they cross along the line of sight (LOS), such as the integrated Sachs–Wolfe effect (ISW) effect \citep{1967ApJ...147...73S} and CMB lensing \citep{1998PhRvD..58b3003Z,1999PhRvL..82.2636S,2000PhRvD..62d3007H,2002ApJ...574..566H,2006PhR...429....1L}.  The two are complementary in  probing the gravitational potential $\phi$ of the universe. ISW probes the gravitational potential decay rate ($DR$) $\dot{\phi}$.  In a flat universe obeying general relativity,   $\dot{\phi}\neq 0$ at late-time only if dark energy exists. This makes it a unique probe of dark energy. Its detection is made possible through the ISW-LSS cross-correlation \citep{1996PhRvL..76..575C,2000ApJ...538...57S,2003ApJ...597L..89F,2004Natur.427...45B,2004PhRvD..70h3536A,2004PhRvD..69h3524A,2004MNRAS.350L..37F,2004ApJ...608...10N,2005NewAR..49...75B,2005PhRvD..72d3525P,2005PhRvD..71l3521C,2006MNRAS.365..891V,2006astro.ph..2398M,2007MNRAS.381.1347C,2007MNRAS.377.1085R,2008MNRAS.386.2161R,2010A&A...520A.101H,2010MNRAS.404..532M,2012MNRAS.427.3044S,2012MNRAS.426.2581G,2014A&A...571A..19P,Shajib_2016,2016A&A...594A..21P,PhysRevD.97.103514,2019MNRAS.484.5267K,2022MNRAS.tmp.2016B}. However, such measurement suffers from degeneracy between  $\dot{\phi}$ and galaxy bias/matter clustering. 

This degeneracy can be broken in a model independent way combining lensing-LSS cross-correlation measurement \citep{2006ApJ...647...55Z}. Essentially, the ratio of two cross-correlations measures $\dot{\phi}/\phi$,\footnote{For more general cases such as the case of modified gravity or dark energy with significant anisotropic stress,  $\phi$ should be replaced with the lensing potential $\Phi_L\equiv  (\phi-\psi)/2$. $\phi$ and $\psi$ are defined through  $d\tau^2=(1+2\psi)dt^2-(1+2\phi)\gamma_{ij}dx^idx^j$ in the Conformal Newtonian Gauge.} up to a prefactor depending on the geometry of the universe but free of galaxy bias and matter clustering. Furthermore, it is less sensitive to the survey masks, whose impacts on the cross-correlation measurements are largely canceled out  in the ratio.  In \cite{2021MNRAS.500.3838D,2021ApJ...923..153D}, we have measured both cross-correlations combining DESI imaging surveys and Planck maps. We explored two LSS tracers, namely the traditional galaxy positions, and low-density points (LDP) recently proposed by \cite{2019ApJ...874....7D}. The ISW effect  was detected at 3.2$\sigma$ and the CMB lensing was detected at 56$\sigma$. Here we combine these  measurements to determine the decay rate. Due to measurement errors, we can not simply take the ratio, otherwise the measured ratio will be biased. We have developed a unbiased estimator of measuring the ratio of two data sets.  (Sun et al., in preparation) and will apply it here.  There are also other studies that using the Planck CMB lensing map to help calibrate the bias of the sources in the ISW measurements, but in different ways \citep{2015PhRvD..91h3533F,2016A&A...594A..21P}. 

This paper is organized as follows. \S\ref{sec:data}  introduces the methodology and data sets used in our analysis. \S\ref{sec:result} presents the measured decay rates at three redshifts. They significantly improve the dark energy constraint over existing baryon acoustic oscillation (BAO) or type Ia supernovae measurements. We discuss and conclude in \S\ref{sec:conclusion}. The estimator adopted to measure $DR$ is briefly introduced in the Appendix \ref{sec:likelihood}. 
\begin{figure}[!htb]
    \centering
     \subfigure{
     \includegraphics[width=1\linewidth, clip]{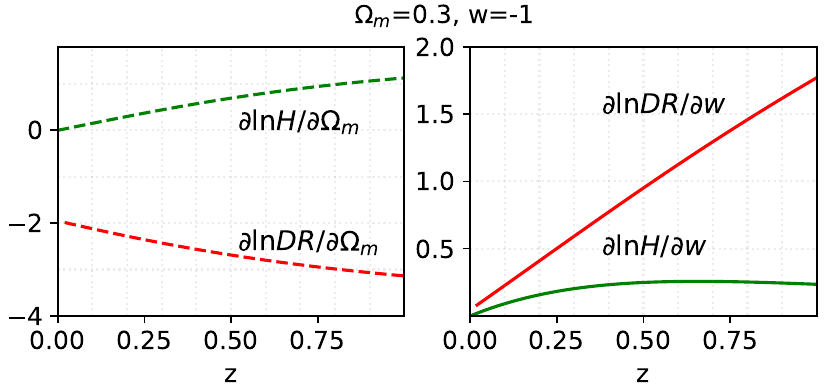}}
   \caption{The sensitivities of $DR$ to $\Omega_m$ and $w$, quantified by $\partial {\rm ln} DR/\partial \Omega_m$ (left panel) and $\partial {\rm ln} DR/\partial w$ (right panel). The results for $H(z)$  are plotted for comparison. For identical fractional measurement error, the constraint of $\Omega_m$ from $DR$ will be a factor of 4 better than that from $H$ on the median, and constraint of $w$ from $DR$ will be a factor of 7 better than that from $H$ for $0<z<1$. A cosmology of $\Omega_m=0.3$ and $w=-1$ is used for the plots. } 
    \label{fig:c-DR}
\end{figure}
\section{Methodology and Data}
\label{sec:data}
\subsection{Isolating the Decay Rate ($DR$) of Gravitational Potential}
\cite{2006ApJ...647...55Z} pointed out a proportionality relation between the ISW-LSS cross-power spectrum $C_{Ig}$ and lensing-LSS cross-power spectrum $C_{\phi g}$, 
\begin{equation}
\label{eq:DR}
C_{Ig}(\ell)\simeq DR(z_m)C_{\phi g}(\ell)\ .
\end{equation}
$z_m$ is the mean redshift of a chosen galaxy redshift bin, namely galaxies located at $z_m-\Delta z/2<z<z_m+\Delta z/2$. The coefficient $DR$ is the decay rate that we can measure, 
\begin{equation}
\label{eq:DRz}
    DR(z)=\left(-\frac{d\ln D_\phi}{d\ln a}\right)\left( \frac{aH(z)/c}{W_L(z)}\right)\ .
\end{equation}
This relations holds for both flat and curved universes. For brevity we focus on a flat universe hereafter. 
$H(z)$, $a$ and $c$ are  the Hubble parameter at redshift $z$, the scale factor and the speed of light respectively. $D_\phi$ is the linear growth factor of gravitational potential $\phi$ \citep{2005PhRvD..72d3529L}. Along with the onset of dark energy, $\phi$ decays with time. So in Eq. \ref{eq:DRz} we include explicitly a negative sign to make $DR$ positive.  Notice that $DR$ defined above differs from the desired decay rate $d\ln D_\phi/d\ln a$ by factors in the last parenthesis, since lensing has an extra geometry dependence $W_L(z)=[1-\chi(z)/\chi(z_s)]/\chi(z)$. $\chi(z)=\int cdz/H(z)$ is the comoving radial coordinate. For CMB lensing, the source redshift  $z_s=1100$. Fortunately,  these extra factors do not depend on $H_0$ and thus avoid uncertainties in $H_0$. 

The above relation is expected to be valid for $z_m\gtrsim0.2$ and $z_m>\Delta z$ \citep{2006ApJ...647...55Z}. We numerically verify that it is accurate to $5\%$ over the multipole range $\ell<300$.\footnote{The accuracy can be further improved when necessary. For example, we may replace $z_m$ by the average redshift of galaxies. } This is sufficiently precise given the $\sim 30\%$ measurement error in $DR$. 

At redshifts where the ISW effect can be detected ($z\la 1$), $DR$ is determined by the matter density $\Omega_m$ and the dark energy equation of state $w$. 
Due to the intrinsic dependence of decay rate on $\dot{\phi}/\phi$, its sensitivity to $w$ is a factor of $3/4/5$ higher at $z=0.3/0.5/0.7$ than that of $H$ (Fig. \ref{fig:c-DR}). This superior sensitivity also holds for $\Omega_m$ (Fig. \ref{fig:c-DR}). Later on we will find that such superior sensitivity partly compensates its significantly weaker measurement S/N, and leads to significant improvement in DE constraint over BAO and SNe Ia.

\begin{table}[]
    \centering
    \begin{tabular}{cccc}
    \hline
Galaxy redshift & $N_{\rm g}$  & $DR$(LDP) &$DR$(gal)\\
\hline
$0.2<z^P<0.4$ & $1.33\times 10^6$ & $0.094\pm0.058$ & $0.087\pm0.054$ \\
$0.4<z^P<0.6$ & $1.66\times 10^6$ & $0.180\pm0.068$ & $0.176\pm0.065$\\
$0.6<z^P<0.8$ & $1.54\times 10^6$ & $0.057\pm0.069$ & $0.043\pm0.068$\\
\hline
    \end{tabular}
    \caption{The galaxy samples selected from DESI imaging survey DR8 and the measured gravitational potential decay rate $DR$. We use both LDPs (low-density points) and galaxy positions as LSS tracers. }
    \label{tab:DR}
\end{table}

\begin{figure*}
    \centering
     \subfigure{
     \includegraphics[width=1\linewidth, clip]{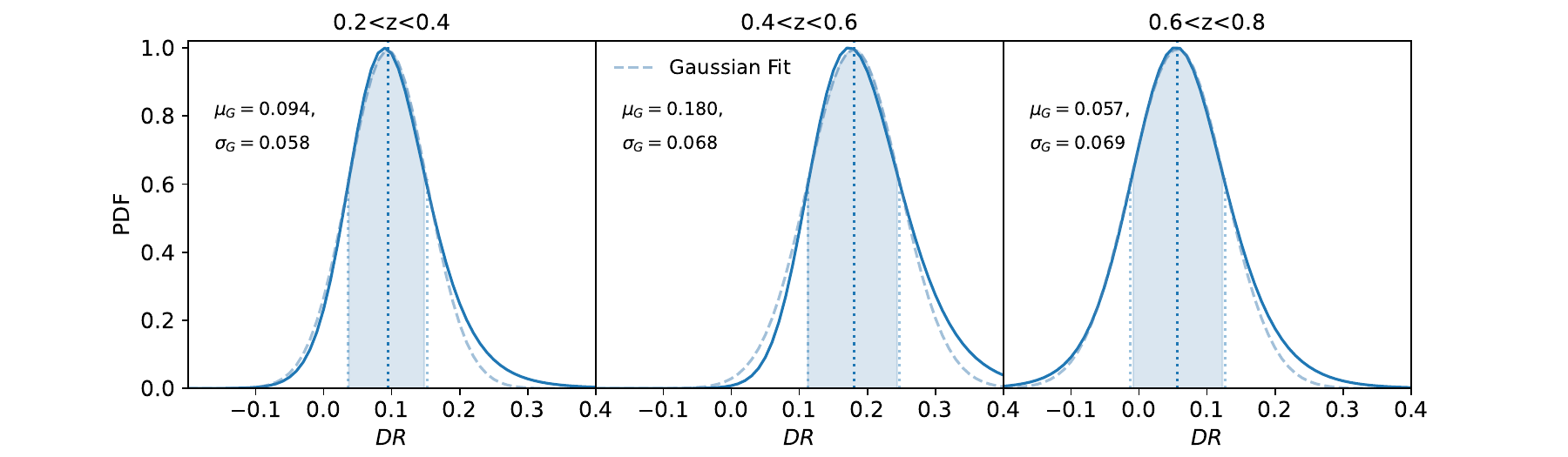}}
   \caption{The probability distribution function of $DR$ (blue solid line) obtained by the  estimator of Eq. \ref{eq:likelihood}. It is well described by a Gaussian function (blue dashed line), for which we list the best-fitting values of the PDF. The PDF is normalized with its peak amplitude.} 
    \label{fig:pdf_3DR}
\end{figure*}
\subsection{CMB Temperature Map and Lensing Map}
In the following, we use $C_{I l}$ and $C_{\phi l}$ to represent the ISW-LDP cross-power spectrum and lensing-LDP cross-power spectrum. 
We use the Planck SMICA temperature product \citep{2016A&A...594A..11P} for the ISW-LDP cross-correlation measurement and the lensing product \citep{2020A&A...641A...8P} for the lensing potential-LDP cross-correlation measurement. For both products, the thermal Sunyaev-Zeldovich (tSZ) effect has been deprojected. The Planck temperature product is provided as a full-sky map, while the lensing product is provided as spherical harmonic coefficients $a_{lm}$ of the lensing convergence in HEALPix FITS format \citep{2005ApJ...622..759G} with $\ell_{max}$ = 4096, for which we multiply them by a factor of  $2/\ell/(\ell+1)$ to produce the lensing potential.

To match the resolution of the LSS tracer density map, we downgrade the above CMB products to a lower resolution with Nside = 512. During the analysis, we adopt the Planck mask to remove pixels contaminated by Galactic dust or known points sources.  Notice that to avoid systematic biases to the lensing data introduced by the mask effect,  aliasing effect and reconstruction noise, we apply a tophat cut in multipole space for generating the lensing potential for which the details are introduced in Appendix \ref{sec:filter}. The same filter is applied to the CMB temperature to eliminate its effect on the DR measurement.

\subsection{Galaxy Catalogue and LDP Identification}
\label{sec:galaxy}
For LSS tracers, we base on the photo-z galaxy catalog from the DESI Legacy Imaging Surveys. The DESI instrument is designed to measure the redshifts of galaxies and quasars from the northern hemisphere in a 14,000 ${\rm deg}^2$ survey, for which the imaging data are provided by three projects: {\it the Beijing-Arizona Sky Survey} (BASS), {\it the Mayall z-band Legacy Survey} (MzLS), {\it the DECam Legacy Survey} (DECaLS). 
In combination with the {\it Dark Energy Survey} (DES), the joint sky coverage of the Data Release 8 (DR8) of the Legacy Surveys  approaches $\sim$ 20000 ${\rm deg}^2$, which is a key to reduce the statistical error in the ISW measurement. We select galaxies from the DR8 photometric galaxy catalog \footnote{\url{http://batc.bao.ac.cn/~zouhu/doku.php?id=projects:desi_photoz:;}} and form volume-limited galaxy samples. The DR8 photometric catalog \citep{2019ApJS..242....8Z} is selected upon the three optical bands ($g$, $r$, $z$) and mid-infrared bands observed by the Wide-field Infrared Survey Explorer satellite~\citep{2019ApJS..242....8Z,2016AAS...22831702S,2005IJMPA..20.3121F,2016AAS...22831701B,2018ApJS..239...18A}. The photo-z of each galaxy is estimated by a local linear regression algorithm \citep{2016MNRAS.460.1371B,2018ApJ...862...12G}.
The catalogue also provides apparent magnitudes in g, r, z bands, and stellar masses of galaxies.

 LDPs depend on the galaxy sample selected. Following \citet{2021MNRAS.500.3838D}, we select galaxies with r-band absolute magnitudes\footnote{The absolute magnitudes used have not been K-corrected. Since galaxies within the same photo-z bin have similar K-correction and the LDP generation is sensitive only to relative brightness between these galaxies, this lack of K- correction is not an issue for our purpose.} $M_r<-21.5$ and $0.2<\leq z^P<0.4$.  This results into $1.33\times 10^6$ galaxies.  We exclude all regions within $R_s=3^{'}$ radius \footnote{Larger Rs or more galaxies locates denser low density areas but less number of LDPs. The adopted $R_s=3^{'}$ is a balance between the two.} of any galaxies in this sample and define the remaining sky positions as LDPs. Statistically speaking, LDPs  correspond to underdense regions \citep{2019ApJ...874....7D,2021MNRAS.500.3838D}. 
We apply the same operations on galaxies at $0.4<z^P<0.6$ and $0.6<z^P<0.8$ respectively (Table \ref{tab:DR}).

We sample LDPs on equal-area HEALPix grids at $N_{\rm side}=4096$, which corresponds to an angular resolution of $0.859'$. We then follow \cite{2021MNRAS.500.3838D}  to define the pixelized LDP over-density field $\delta_{\rm LDP}$ at Nside=512 resolution,  
\begin{equation}
\label{eq:fldp}
\delta_{\rm LDP}=\frac{f_{\rm LDP}-\overline{f}_{\rm LDP}}{\overline{f}_{\rm LDP}} \ .
\end{equation}
Here $f_{\rm LDP}$ is the fraction of each $N_{\rm side}=512$ pixel occupied by LDPs, and $\bar{f}_{\rm LDP}$ is the mean quantity averaged over the survey area.  $\delta_{\rm LDP}$ is tightly correlated with the matter overdensity $\delta_m$ and galaxy number overdensity $\delta_g$, as we have verified in N-body simulations. The adopted magnitude cut, $\Delta z$, $R_s$ and the definition of $\delta_{\rm LDP}$ have yielded significant detection of the ISW-LDP cross-correlation \citep{2021MNRAS.500.3838D}.

A combined mask from both galaxy catalogue\footnote{The survey mask is generated from the available random catalogs provided by the DR8 website, which contain the number of observation in the g, r, z bands and a general purpose artifact flag MASKBITS, according to the coordinates drawn from the observed distribution. We refer readers to \cite{2021ApJ...923..153D} for the detailed procedure.} and the Planck survey is applied to the LDP over-density field. Furthermore, to avoid possible foreground contaminations from the Galactic plane, we put an additional galactic mask with $|\rm{DEC}|>30^\circ$ to both the CMB maps and the LDP (galaxy) maps. We perform all the analysis in the Galactic coordinate system.

\section{$DR$ measurements and improved dark energy constraints}
\label{sec:result}

\begin{figure}
    \centering
    \subfigure{
     \includegraphics[width=1\linewidth, clip]{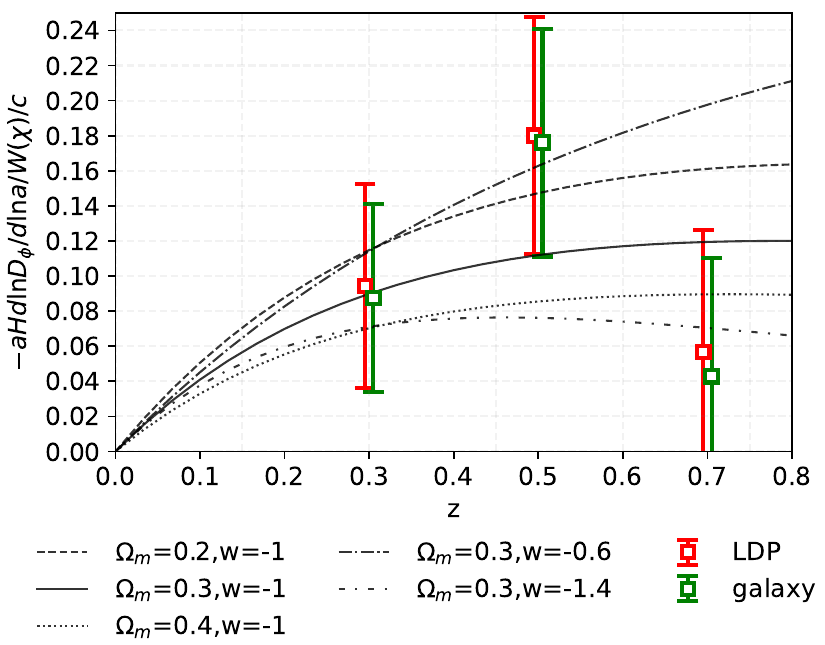}}
   \caption{The gravitational potential decay rate $DR$ measured at $0.2<z<0.4$, $0.4<z<0.6$ and $0.6<z<0.8$.  Red(green) data points show the results obtained with LDPs(galaxies).  Theoretical curves of five $w$CDM cosmologies are also shown for demonstration.} 
    \label{fig:DR}
\end{figure}

\subsection{Measurement of  DR}
The selection criterion in Section \ref{sec:galaxy} results in a total number of $\sim 4.5\times10^6$ galaxies in three  redshift bins ($[0.2, 0.4]$, $[0.4, 0.6]$ \& $[0.6, 0.8]$, Table \ref{tab:DR}). We then follow the same procedure in \citet{2021MNRAS.500.3838D,2021ApJ...923..153D} to measure the ISW-LDP cross-power spectrum $C_{Il}$ and the CMB lensing-LDP cross-power spectrum $C_{\phi l}$. Considering that $C_{Il}$ mainly arises from large angular scale ($\ell<50$) while the measurement of $C_{\phi l}$ is noisy at $\ell<10$, we choose $(\ell_{min}, \ell_{max})\sim (9,117)$ and six  angular bins equally spaced in $\ln \ell$.

Instead of directly taking the ratio of these two cross-correlations, we use the likelihood method for estimating $DR$ and the corresponding error bar (Appendix \ref{sec:likelihood}). It directly evaluates $P(DR)$ using the exact analytical expression (Eq. \ref{eq:likelihood}). Since it involves no multi-parameter fitting, it is computationally fast. Furthermore, the resulting $P(DR)$ is unbiased and relies on no cosmological assumptions other than the proportionality relation. The obtained $P(DR)$ of three redshift bins are shown in Fig. \ref{fig:pdf_3DR}. $P(DR)$ is very close to Gaussian. Therefore later we will adopt a Gaussian $P(DR)$ in cosmological parameter fitting. 

The results of $DR$ are shown in Table \ref{tab:DR} \&  Fig.\ref{fig:DR}. The detection significances of three redshifts are $1.62\sigma$, $2.66\sigma$ and $0.82\sigma$, respectively. The total significance is $3.2\sigma$. For comparison, we also measure DR with the same set of galaxies used for generating LDPs, and obtain a total $\sim 3.2\sigma$ measurement. The measurements including the S/N are fully consistent with each other (Table.\ref{tab:DR}). Because the sign of the cross-correlations of $C_{Il}$ and $C_{\phi l}$ is flipped for LDP, the sign of DR is identical to that measured with galaxies.

\begin{figure*}[!htb]
    \centering
     \subfigure{
     \includegraphics[width=0.43\linewidth, clip]{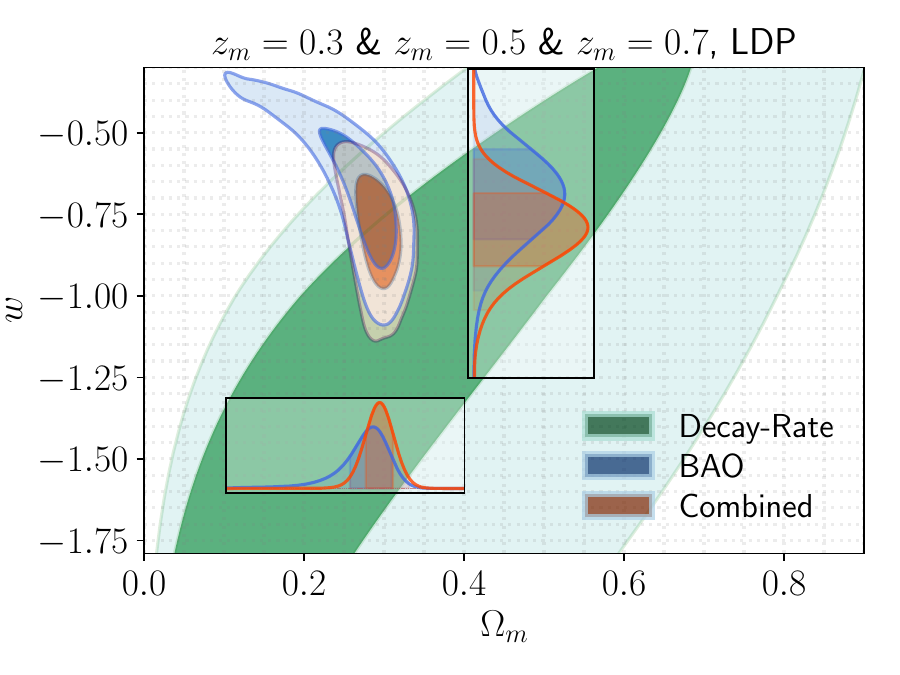}}
     \subfigure{
     \includegraphics[width=0.43\linewidth, clip]{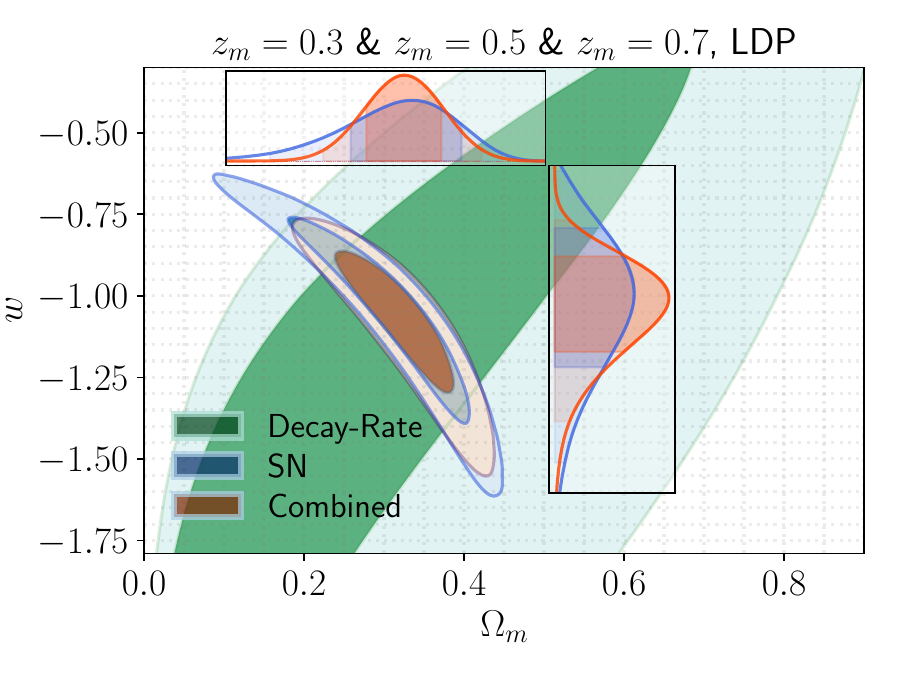}}
   \caption{Constraints on the flat $w$CDM model, from BAO/DR/BAO+DR (left panel) and SN/DR/SN+DR (right panel). Partly due to the $\Omega_m$-$w$ degeneracy direction and partly due to its higher sensitivity on $\Omega_m$ and $w$,  the inclusion of DR significantly improves over BAO or SN alone.  
   }
    \label{fig:3z-contour}
\end{figure*}

\begin{figure}[!htb]
    \centering
     \subfigure{
     \includegraphics[width=1\linewidth, clip]{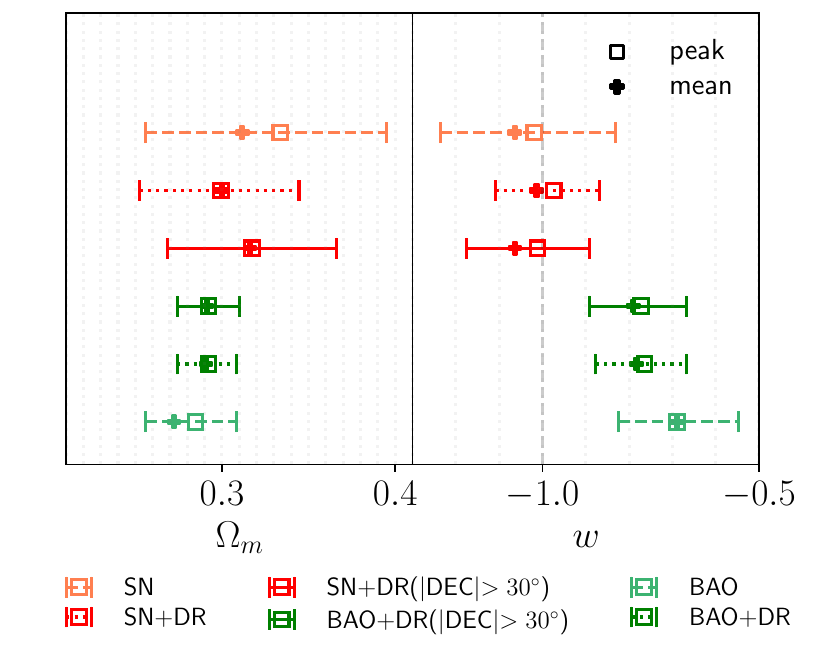}}
   \caption{Improvement in $w$CMD constraints by including the $DR$ measurement. ``peak'' refers to the best-fit value of the parameter and ``mean" refers to the average value of the parameter. The improvement in parameter uncertainty is $1.4$-$2$. $\left|{\rm DEC}>30^\circ\right|$ is an additional mask adopted to all maps used for deriving the main results in this paper (solid lines). We also show the results obtained without adopting this mask for comparison (dotted lines), which are consistent with the solid lines. }
    \label{fig:constraint}
\end{figure}

\subsection{Constraint on a Flat $w$CDM Model}
For the dark energy constraint, we restrict the analysis to the flat $w$CDM cosmology. 
To demonstrate its constraining power on DE, we show $DR(z)$ of various flat $w$CDM cosmology in Fig. \ref{fig:DR}.   $DR$ first increases with redshift ($DR\propto azH$) until $z\sim 1$, and then begins to decrease due to vanishing $\dot{\phi}$.  A lower $\Omega_m$ or a less negative $w$ leads to a higher $DR$.  In particular, since $DR=0$ if $\Omega_m=1$, a non-zero measurement provides a smoking gun evidence of dark energy.

Nevertheless, the $\sim 3\sigma$ measurement of $DR$ seems disappointing in constraining power in $w$. However, there are three points that boost its DE constraining power to be much stronger than than that implied by the measurement S/N.  We emphasize two here and postpone the third after the likelihood analysis. Firstly, as shown in Fig. \ref{fig:c-DR}, the sensitivity of $DR$ to $w$ and $\Omega_m$ is a factor of $\sim 4$ higher than that of $H$. Secondly, $DR$ relies on fewer parameters than other data. For flat $\Lambda$CDM, it only relies on $\Omega_m$. For flat $w$CDM, it  relies on $\Omega_m$ and $w$. As comparisons, both BAO and SNe Ia depend on extra cosmological parameters such as $H_0$ (or $r_s$), and extra nuisance parameters in the data analysis.

Since $p(DR)$ is nearly Gaussian, we estimate the posterior distribution of $(\Omega_m,w)$ by $\mathcal{L}\propto \exp(-\chi^2/2)$ and 
\begin{equation}
\chi^2=\sum_{i=1}^3\frac{(DR_{\rm obs}(z_i)-DR_{\rm model}(\Omega_m,w,z_i))^2}{\sigma^2_{DR}(z_i)}\ .
\end{equation}
Here  $\sigma_{DR}(z_i)$  is the corresponding error bar of the $i$-th $DR$ measurement.

Fig.\ref{fig:3z-contour} shows the constraints from $DR$ measured by LDPs.\footnote{ Constraints with galaxy positions are very similar. Therefore we will not show them here.} Due to limited S/N, $DR$ measurements alone suffer from a strong degeneracy between $\Omega_m$ and $w$. Since for a flat universe $\Omega_{\rm DE}=1-\Omega_m$, the degeneracy direction just restates that higher dark energy density or less negative $w$ (and therefore longer duration of dark energy dominance) results into higher $DR$. This behavior is consistent with behaviors of theoretical curves in Fig. \ref{fig:DR}.

\begin{table*}
\centering
\caption{Marginalized values and 68$\%$ confidence limits of ($\Omega_m$, w) estimated from DR when in combination with different probes and with different choices of redshift. \label{Table2}}
\begin{minipage}{180mm}
\centering
\begin{tabular}{cc|ccc|ccc}
\hline
\hline
&&$\Omega_m$ (best-fit ),  & $\langle\Omega_m\rangle$, & $\sigma(\Omega_m)$ &$w$, & $\langle w \rangle$, & $\sigma(w)$\\

\hline

                                          &BAO     & $0.285_{-0.030}^{+0.024}$,&0.272, &0.035 &$-0.682_{-0.14}^{+0.13}$,&-0.69, &0.144\\
 LDP($\left|{\rm DEC}\right|>30^\circ$)    &BAO+DR  & $0.292_{-0.018}^{+0.018}$,&0.291, &0.019 &$-0.772_{-0.120}^{+0.105}$,&-0.79, &0.118\\
 LDP($\left|{\rm DEC}\right|>30^\circ$)    &BAO+DR (joint 3z)  & $0.290_{-0.018}^{+0.018}$,&0.289, &0.020 &$-0.756_{-0.120}^{+0.105}$,&-0.773, &0.119\\
LDP                                       &BAO+DR  & $0.292_{-0.018}^{+0.016}$,&0.291, &0.019 &$-0.764_{-0.112}^{+0.097}$,&-0.783, &0.112\\
galaxy($\left|{\rm DEC}\right|>30^\circ$) &BAO+DR  & $0.294_{-0.018}^{+0.016}$,&0.293, &0.019 &$-0.787_{-0.120}^{+0.105}$,&-0.807, &0.117\\
galaxy($\left|{\rm DEC}\right|>30^\circ$) &BAO+DR (joint 3z)  & $0.292_{-0.018}^{+0.016}$,&0.292, &0.019 &$-0.779_{-0.120}^{+0.105}$,&-0.799, &0.118\\
galaxy                                    &BAO+DR  & $0.294_{-0.018}^{+0.016}$,&0.293, &0.018 &$-0.779_{-0.120}^{+0.097}$,&-0.80, &0.112\\
\hline
                                          &SN    & $0.334_{-0.078}^{+0.061}$,&0.312, &0.071 &$-1.019_{-0.217}^{+0.187}$,&-1.06, &0.21\\
LDP($\left|{\rm DEC}\right|>30^\circ$)    &SN+DR & $0.317_{-0.049}^{+0.049}$,&0.316, &0.049 &$-1.011_{-0.164}^{+0.120}$,&-1.064, &0.148\\
LDP($\left|{\rm DEC}\right|>30^\circ$)    &SN+DR (jont 3z) & $0.310_{-0.051}^{+0.051}$,&0.309, &0.051 &$-0.996_{-0.150}^{+0.120}$,&-1.042, &0.148\\
LDP                                       &SN+DR & $0.299_{-0.047}^{+0.045}$,&0.30,  &0.046 &$-0.974_{-0.135}^{+0.105}$,&-1.013, &0.126\\
galaxy($\left|{\rm DEC}\right|>30^\circ$) &SN+DR & $0.325_{-0.047}^{+0.047}$,&0.323, &0.048 &$-1.034_{-0.165}^{+0.130}$,&-1.084, &0.149\\
galaxy($\left|{\rm DEC}\right|>30^\circ$) &SN+DR (joint 3z) & $0.321_{-0.047}^{+0.049}$,&0.321, &0.049 &$-1.026_{-0.165}^{+0.120}$,&-1.077, &0.150\\
galaxy                                    &SN+DR & $0.307_{-0.043}^{+0.045}$,&0.308, &0.045 &$-0.996_{-0.134}^{+0.105}$,&-1.037, &0.128\\
\hline
\end{tabular}
\begin{tablenotes}
      \small
      \item Note. --- Columns for $\Omega_m$, $\langle\Omega_m\rangle$ and $\sigma(\Omega_m)$ refers to the best-fit value of the matter density, the average value and the scatter with respect to $\langle\Omega_m\rangle$. Columns of $w$ have the same meaning. We show the results obtained by adopting an additional galactic mask of $|\rm{DEC}|>30^\circ$ for measuring DR or not. The former result is more safe to use, while the latter shows smaller $\sigma_w$ (especially for DR+SN). We also show the results obtained from a joint analysis of DR of three redshift slices for consistency check, labelled as `joint 3z'. These results are consistent with each other.
    \end{tablenotes}
\end{minipage}
\end{table*}

Nevertheless, $\Omega_m$-$w$ constraints from $DR$ are still useful, since they are highly complementary to constraints from BAO and SNe Ia. Fig.\ref{fig:3z-contour} shows the contours from SDSS BAO data \citep{2021PhRvD.103h3533A},\footnote{We use the MCMC chains and likelihoods from the https://www.sdss.org/science/cosmology-results-from-eboss.} which includes observations of galaxy and quasar samples from the SDSS, BOSS, and the eBOSS surveys at $z<2.2$ and Ly$\alpha$ forest observations over $2<z<3.5$ \citep{2020MNRAS.499..210N,2021MNRAS.500.1201H,2021MNRAS.501.5616D,2020MNRAS.499.5527T,2021MNRAS.500.3254R,2020MNRAS.498.2492G,2021MNRAS.500..736B}. It also shows the contours from the Pantheon SNe Ia data \citep{2018ApJ...859..101S}. Clearly,  the $\Omega_m$-$w$ degeneracy direction of $DR$ is largely orthogonal to those of BAO and SN. The reason is that, for a flat universe both BAO and SN constrain $\Omega_m$ and $w$ through their impacts on the same $H(z)$, although with different redshift weights. So the degeneacy direction is $\delta w\simeq c \delta \Omega_m$.  $c\sim -(\partial  H/\partial \Omega_m)/(\partial H/\partial w)_{z_*}=-[(1+z_*)^{-3w}-1]/[3(1-\Omega_m)\ln(1+z_*)]$. $z_*$ is the redshift where most constraining power locates. Therefore for both BAO and SN, $c_{\rm BAO}<0$ and $c_{\rm SN}<0$. In contrast, $c_{\rm DR}>0$ (Fig.\ref{fig:c-DR}).

Therefore the inclusion of DR measurement significantly improves over BAO/SN constraints of $\Omega_m$ and $w$ (Table \ref{Table2} \& Fig.\ref{fig:constraint}). Comparing to SDSS BAO alone,  DR+BAO reduces $\sigma_{\Omega_m}$ (the statistical error in $\Omega_m$) by a factor of 1.8 and $\sigma_w$ by a factor of  1.29. For DR+SN, the factors are 1.45 \&  1.42 respectively.
We also find that the value of $\Omega_m$ estimated from BAO$+$DR ($0.294_{-0.018}^{+0.018}$) is more consistent with that of SN+DR ($0.317_{-0.049}^{+0.049}$) comparing to the case of BAO only ($0.285_{-0.03}^{+0.024}$) and SN ($0.334_{-0.078}^{+0.061}$) only. Another interesting finding is that the inclusion of $DR$ alleviates the slight tension between $w$ constrained form BAO/SN. Comparing to BAO, DR+BAO shift the bestfit $w$ closer to $-1$. As a cross-check, we also perform the analysis with DR measured from galaxy positions. The obtained  constraints are consistent with LDP (Table \ref{Table2}).

\section{Discussion}
\label{sec:conclusion}
We detect the gravitational potential decay rate ($DR$) in a model-independent way by combining the ISW-LSS cross-correlation with CMB lensing-LSS cross-correlation. The large overlap in survey region between DESI DR8 photo-z catalogue and Planck 2018 CMB data release enables us to detect $DR$ at a total significance of 3.2$\sigma$ within the redshift range of $0.2<z<0.8$. We demonstrate that $DR$ highly complements BAO/SN, and therefore significantly improves their $w$CDM constraints. 

In the above, we measure DR independently for each redshift. One may wonder whether a joint analysis of the cross-power spectrum from three redshift slices could improve the detection of DR, since non-zero cross-correlations are found between different redshifts, possibly caused by the same adopted mask, mixture of galaxies due to the photo-z error, as well as the wide ISW and CMB-lensing kernels. To do so, we consider the full covariance by including autocorrelations, $C_{Il}(z_i)-C_{Il}(z_i)$ ($C_{\phi l}(z_i)-C_{\phi l}(z_i)$), and cross-correlations, $C_{Il}(z_i)-C_{Il}(z_j)$ ($C_{\phi l}(z_i)-C_{\phi l}(z_j)$, $i\neq j$). We then fit values of DR of three redshifts simultaneously (Table \ref{tab:DR-joint3z}). The cross-correlation between different redshifts is weak, so it is expected that values of DR are consistent with the measured results from Table \ref{tab:DR}. Notice that there are only 300 noise FFP10 simulations \citep{2020A&A...641A...8P} available for estimating the full covariance matrix of CMB lensing. While the precision joint measurement of DR requires many more mocks. So the analysis here is just for consistency check.

\begin{table}[]
    \centering
    \begin{tabular}{cccc}
    \hline
Galaxy redshift & $N_{\rm g}$  & $DR$(LDP) &$DR$(gal)\\
\hline
$0.2<z^P<0.4$ & $1.33\times 10^6$ & $0.111\pm0.062$ & $0.097\pm0.056$ \\
$0.4<z^P<0.6$ & $1.66\times 10^6$ & $0.177\pm0.070$ & $0.172\pm0.067$\\
$0.6<z^P<0.8$ & $1.54\times 10^6$ & $0.068\pm0.073$ & $0.046\pm0.070$\\
\hline
    \end{tabular}
    \caption{Similar to Table.\ref{tab:DR}, but with DR obtained by taking into account of the cross-correlations between different redshifts.}
    \label{tab:DR-joint3z}
\end{table}

As the ISW effect peaks around $z\sim0.5$, our current measurements of $DR$ focus on the low redshift ($0.2<z<0.8$). Considering that the optimistic S/N of the full-sky ISW signal for a standard $\Lambda$CDM model is around 7$\sigma$ \citep{2008PhRvD..77l3520G}, our detections of $DR$ are already close to the statistical limit set by limited sky coverage and redshift range.  However, it is necessary to include higher redshift in the future for the following reasons. First, more independent volumes are available for higher z, meaning lower statistical error. Second, the measurement of $DR$ at high z, even a null detection,  is useful for distinguishing between different dark energy models. DESI, with over 30 million spectroscopic galaxy and quasar redshifts out to $z=3.5$ and 14000 deg$^2$ sky coverage, will further improve the DR S/N, while reducing possible systematic errors in the current measurement (e.g. photometric redshift bias).

In addition to the DESI surveys in the northern sky, the {\it Large Synoptic Survey Telescope} (LSST, \citet{2012arXiv1211.0310L}) under construction will cover about 18,000 ${\rm deg}^2$ of the southern sky in the main survey for the next decade, with deeper image depths ($r\sim$ 24.5). This will double the sky coverage of ISW/lensing-LSS cross-correlation measurement and improve DR measurement by another $\sim 40\%$.  

At last, there are still other possibilities to further explore. For example, $DR$ could also be detected through  ISW-LSS cross-correlation $+$ galaxy shear-LSS cross-correlation. Therefore, it can serve as a cross-check for the $DR$ measurement here and help eliminate possible systematics from each other.


\section*{Acknowledgments}
FD is supported by a KIAS  Individual Grant PG079001 at Korea Institute for Advanced Study.
CP is supported by a KIAS Individual Grant PG016904 at Korea Institute for Advanced Study. 
PZ and ZS are supported by National Science Foundation of China (11621303 \& 11653003), the National Key R$\&$D Program of China (2020YFC2201602), NO. CMS-CSST-2021-A02 and  the 111 project.
 FD acknowledges the Korea Institute for Advanced Study for providing computing resources (KIAS Center for Advanced Computation) for this work. The authors thank the HEALPix/healpy software package \citep{Zonca2019,2005ApJ...622..759G}.

\bibliography{dr}
\appendix 
\section{An estimator to  measure $DR$}
\label{sec:likelihood}
If two data vectors (${\bf d}_{1,2}$) are expected to obey ${\bf d}^{\rm theory}_2=R{\bf d}^{\rm theory}_1$, the ratio $R$ can be estimated by Bayesian analysis. To be model independent, we take the parameters to be marginalized as $\lambda\equiv {\bf d}^{\rm theory}_1$ and adopt a flat prior $P(\lambda)$. The marginalization can be done analytically and the the posterior distribution of $R$ is given by 
\begin{equation}
\label{eq:likelihood}
P(R|d_1,d_2)\propto\int P(d_1,d_2|R,\lambda)d\lambda
\propto {\rm det}^{1/2}{\bf Q} \exp\left[\frac{1}{2}(\sum_i {\bf A}_i^T{\bf C}_i^{-1}{\bf d}_i)^T {\bf Q}^{-1} (\sum_i {\bf A}_i^T{\bf C}_i^{-1}{\bf d}_i)^T\right]\ .
\end{equation}
Here ${\bf Q}=\sum_i{\bf A}_i^T{\bf C}^{-1}_i{\bf A}_i$, ${\bf A}_1={\bf I}$, ${\bf A}_2=R{\bf I}$, and ${\bf C}_i$ is the covariance matrix of ${\bf d}_i$. To measure $DR$ by this estimator, we replace  ${\bf d}_1$ and ${\bf d}_2$ with the CMB lensing-LSS cross-correlation $C_{\Phi l}(\ell)$ and the ISW-LSS cross-correlation $C_{{\rm I}l}(\ell)$. We use simulations to determine the covariance matrix. Noise  Planck  Full  Focal  Plane (FFP106) simulations \citep{2020A&A...641A...8P}  are used to determine the mean-field bias and the covariance matrix estimation of $C_{\Phi l}(\ell)$ \citep{2021ApJ...923..153D}.
While for $C_{{\rm I}l}$, we simulate CMB temperature maps with the Planck cosmology to estimate the covariance matrix \citep{2021MNRAS.500.3838D}.  Fig.\ref{fig:pdf_3DR} shows $P(DR)$. We then find the bestfit $DR$ and $\sigma_{\rm DR}$. 

The estimator can be extended to more general cases of the theory-data mapping matrixes ${\bf A}_{1,2}$ and we refer readers to Sun et al. 2022 (in preparation) for details.

\begin{figure}[!htb]
    \centering
     \subfigure{
     \includegraphics[width=0.5\linewidth, clip]{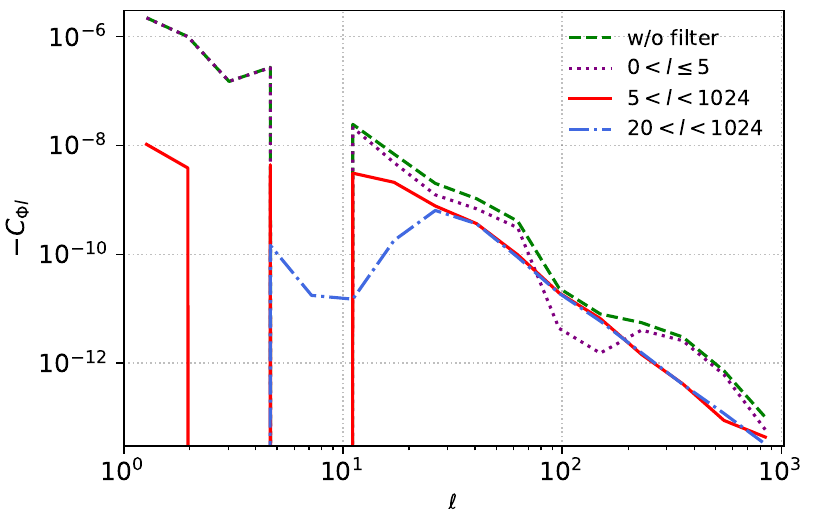}}
   \caption{The lensing potential-LDP cross-correlation measured with different filters upon the masked CMB lensing potential. Green dashed line: without applying any filter to the lensing potential. Red solid line: applying a top hat filter of $5<\ell<1024$ to the lensing potential $\Phi_{\ell m }$. Purple dotted line: applying a top hat filter of $0<\ell\le5$. Blue dashed-dotted line: applying a top hat filter of $20<\ell<1024$.} 
    \label{fig:appen}
\end{figure}
\section{Filtering the CMB data}
\label{sec:filter}
In this work, we produce the CMB lensing potential based on the Planck convergence product $\kappa_{\ell m}$: $\Phi_{\ell m}=2/\ell/(\ell+1)\kappa_{\ell m}$. However, the construction of $\kappa_{\ell m}$ in observation is contaminated with noise, which propagates onto $\Phi_{\ell m}$ and is amplified by a factor of $2/\ell/(\ell+1)$. Therefore, there is a risk for directly using $\Phi_{\ell m}$, especially in the existence of mask. We find that our adopted mask is highly correlated with the lowest multipoles of $\Phi_{\ell m}$ ($\ell\le5$). Consequently, the contaminated cross power spectra at $\ell\le5$ are leaked onto higher $\ell$ and severely interferes our measurement of $C_{\phi l}$ at all scales. To avoid such effect, we adopt a tophat filter with $5<\ell<1024$ for generating the lensing potential map in this study. The result is shown in Fig.\ref{fig:appen}, in which we have performed the scale cut test by varying the range of the tophat filter in multipole space: $\ell_{min}<\ell<\ell_{max}$. We find that $\ell_{min}=5$ is a safe choice as the cross-power spectra of higher multipoles is almost identical to the one when adopting a larger $\ell_{min}$. The choice of $\ell_{max}$ is to prevent the aliasing of high-$\ell$ noise power spectrum \footnote{ In  \cite{2021ApJ...923..153D} we found that  downgrading the lensing convergence map $\kappa_{\ell m}$ from its original resolution to a lower resolution one can introduce aliasing of high-$\ell$ noise power spectrum. An appropriate low-pass filter can prevent such effect. This effect is reduced to a low level when using $\Phi_{\ell m}$. However, we still keep this procedure for safety.}.

\end{document}